\documentclass[12pt,tightenlines,preprintnumbers]{revtex4}
\usepackage{amsmath}
\usepackage{amssymb}
\usepackage{mathptmx}
\usepackage{bm}
\usepackage{ifthen}
\usepackage{graphicx}
\graphicspath{{./}{figures/}}
\allowdisplaybreaks

\newcommand{\bbswitch}{1}
\newcommand{\Lx}{\left(}
\newcommand{\Rx}{\right)}
\newcommand{\LB}{\left[}
\newcommand{\RB}{\right]}
\newcommand{\hypgeo}[4]{{\vphantom{F}}_2F_1\Lx{#1},{\,#2};{\,#3};{\,#4}\Rx}
\newcommand{\ghypgeo}[6]{{\vphantom{F}}_3F_2\Lx{#1},{\,#2},{\,#3};{\,#4},{\,#5};{\,#6}\Rx}
\newcommand{\gghypgeo}[8]{{\vphantom{F}}_4F_3\Lx{#1},{\,#2},{\,#3},{\,#4};{\,#5},{\,#6},{\,#7};{\,#8}\Rx}

\newcommand{\ep}{{\varepsilon}}
\newcommand{\dep}{{\epsilon}}
\newcommand{\eqn}[1]{Eq.\,(\ref{#1})}
\newcommand{\eqns}[2]{Eqs.\,(\ref{#1}-\ref{#2})}
\newcommand{\pLi}[3]{\mbox{Li}^{#3}_{#1}\left(#2\right)}
\newcommand{\nLi}[2]{\mbox{Li}_{#1}\left(#2\right)}

\newcommand{\Limuds}[1]{\mbox{Li}_{#1}\left({-u\over s}\right)}

\newcommand{\Sx}[3]{\mathop{\hbox{\rm S}_{#1,#2}}\nolimits\left(#3\right)}
\newcommand{\nnlo}{{NNLO}}
\newcommand{\qcd} {{QCD}}
\newcommand{\qed} {{QED}}
\newcommand{\lhc} {{LHC}}
\newcommand{\MIi}{{I^{(2)}_3(s)}}
\newcommand{\MIii}{{I^{(2)}_3(u)}}
\newcommand{\MIiii}{{I^{(2)}_3(t)}}
\newcommand{\MIiv}{{I^{(2)}_{4;1}(s)}}
\newcommand{\MIv}{{I^{(2)}_{4;2}(s)}}
\newcommand{\MIvi}{{I^{(2)}_{4;2}(u)}}
\newcommand{\MIvii}{{I^{(2)}_{4;2}(t)}}
\newcommand{\MIviii}{{I^{(2)}_{5;1}(s,u,t)}}
\newcommand{\MIix}{{I^{(2)}_{5;1}(u,s,t)}}
\newcommand{\MIx}{{I^{(2)}_{5;1}(s,t,u)}}
\newcommand{\MIxi}{{I^{(2)}_{5;1}(t,s,u)}}
\newcommand{\MIxii}{{I^{(2)}_{5;2}(s,u,t)}}
\newcommand{\MIxiii}{{I^{(2)}_{5;2}(s,t,u)}}
\newcommand{\MIxiv}{{I^{(2)}_{6}(s)}}
\newcommand{\MIxv}{{I^{(2)}_{7;1}(s,u)}}
\newcommand{\MIxvi}{{I^{(2)}_{7;1}(s,t)}}
\newcommand{\MIxvii}{{I^{(2)}_{7;2}(s,u)}}
\newcommand{\MIxviii}{{I^{(2)}_{7;2}(s,t)}}
\newcommand{\bra}[1]{\left\langle #1\right.}
\newcommand{\ket}[1]{\left| #1\right\rangle}

\newcommand{\feynsl}[1]{{
  \setbox0=\hbox{/} \setbox1=\hbox{$#1$}
  \dimen0=\wd0 \advance\dimen0 by -\wd1 \divide\dimen0 by 2
  \ifdim\wd0>\wd1 \raise.15ex\copy0\kern-\wd0\kern\dimen0\copy1\kern\dimen0
  \else \kern-\dimen0\raise.15ex\copy0\kern-\dimen0\kern-\wd1\copy1\fi}}
\newcommand{\aspi} {{\Lx\frac{\alpha_{s}}{\pi}\Rx}}
\newcommand{\aepi} {{\Lx\frac{\alpha}{\pi}\Rx}}
\newcommand{\asbarepi} {{\Lx\frac{\alpha_s^{B}}{\pi}\Rx}}
\newcommand{\aebarepi} {{\Lx\frac{\alpha^{B}}{\pi}\Rx}}
\newcommand{\MSbar}{$\overline{\mbox{MS}}$ }
\newcommand{\logsdmus}{\log\left({s\over\mu^2}\right)}
\newcommand{\logsdmusa}[1]{\log^{#1}\left({s\over\mu^2}\right)}

\newcommand{\logmtds}{\log\left({-t\over s}\right)}
\newcommand{\logmtdsa}[1]{\log^{#1}\left({-t\over s}\right)}
\newcommand{\logmuds}{\log\left({-u\over s}\right)}
\newcommand{\logmudsa}[1]{\log^{#1}\left({-u\over s}\right)}
\newcommand{\z}[1]{\zeta_{#1}}
\newcommand{\CF}{C_F}
\newcommand{\nc}{N_c}
\newcommand{\gammaE}{\gamma_{E}}
\newcommand{\Gammap}[2]{\Gamma^{#2}(#1)}
\newcommand{\Gamman}[1]{\Gamma(#1)}

\newcommand{\nlog}[1]{\log\left(#1\right)}
\let\*\,
\newcommand{\includegraphicsNEW}[3]{
\ifthenelse{\equal{\bbswitch}{1}}{\includegraphics[#1,#2]{#3}}{\includegraphics[#2]{#3}}
                                   }

\begin{document}

\begin{titlepage}

July 24, 2011
\preprint{MPP-2011-90}

\title{Two-Loop Virtual Corrections to Drell-Yan Production at order $\alpha_s\,\alpha^3$}

\author{William~B.~Kilgore}
\affiliation{Physics Department, Brookhaven National Laboratory,
  Upton, New York
  11973, U.S.A.\\
  {\tt [kilgore@bnl.gov]} }
\author{Christian~Sturm}
\affiliation{
   Max-Planck-Institut f\"ur Physik,
   (Werner-Heisenberg-Institut),\\
   F\"ohringer Ring 6, 80805 M\"unchen, Germany\\
  {\tt [sturm@mpp.mpg.de]} }

\begin{abstract}
\vspace*{9cm} 
The Drell-Yan mechanism for the production of lepton pairs is one of
the most basic processes for physics studies at hadron colliders.  It
is therefore important to have accurate theoretical predictions. In
this work we compute the two-loop virtual mixed \qcd$\times$\qed\
corrections to Drell-Yan production. We evaluate the Feynman diagrams
by decomposing the amplitudes into a set of known master integrals and
their coefficients, which allows us to derive an analytical result. We
also perform a detailed study of the ultraviolet and infrared
structure of the two-loop amplitude and the corresponding poles in
$\varepsilon$.
\end{abstract}

\maketitle
\end{titlepage}

\section{Introduction\label{sec:Intro}}
The Drell-Yan process is one of the most precise probes available at
hadron colliders.  It allows for precise measurements of the gauge
boson masses~\cite{Aaltonen:2007ypa,Aaltonen:2007ps,Abazov:2009cp},
widths~\cite{Aaltonen:2007mg,Abazov:2009vs} and
asymmetries~\cite{Abazov:2008xq} and is very sensitive to physics
beyond the Standard Model like new gauge
bosons~\cite{Abazov:2007bs,Aaltonen:2008vx,Aaltonen:2008ah,Abazov:2010ti}.
One reason it is such a powerful probe is its very simplicity.  Its
experimental signature, two leptons plus anything, is quite robust
against radiative emission.  Theoretically, it is perhaps the simplest
process to compute at hadron colliders.  It was the first hadronic
scattering process to be computed at next-to-next-to-leading (\nnlo)
in \qcd~\cite{Hamberg:1991np,Harlander:2002wh}, almost twenty years
ago.

In recent years, it has become clear that electroweak
corrections~\cite{Baur:1997wa,Baur:1998kt,Baur:2001ze,Dittmaier:2001ay}
to Drell-Yan production are also very important.  Electroweak
corrections can distort the line-shape and thereby affect the
measurement of the gauge boson masses.  Radiative corrections can also
become very large at the high energies (several hundred GeV) which
will be probed at the \lhc.

An important next step to refining the prediction for Drell-Yan
production is the calculation of the complete mixed \qcd\ and
electroweak corrections.  Currently, only the virtual corrections to
the quark -- gauge boson vertex are known in the
literature~\cite{Kotikov:2007vr}.  We embark on this project by
computing the simplest gauge-invariant part, the mixed
\qcd$\times$\qed\ virtual corrections.  That is, we ignore all $W$ and
$Z$ boson interactions, and consider only virtual photon and gluon
exchanges.  In addition, we take all fermions to be massless (except
the top quark, which does not enter into this part of the
calculation).  For most of the calculation, there is no barrier to
including a non-vanishing lepton mass.  For the box contributions,
however, one would need new master integrals with two massive external
legs and massive internal propagators.
The effects of non-vanishing masses (at least for components that do not
involve box contributions) will be added at a later stage of the
project.

The outline of this paper is as follows. In Section~\ref{sec:Notation}
we define some generalities and our notation. In Section
~\ref{sec:calculation} we give an outline of the calculation and in
Section~\ref{sec:poles} discuss the structure of the ultraviolet and
infrared poles.  The results are presented in
Section~\ref{sec:results} and we present our conclusions in
Section~\ref{sec:summary}. The Appendix contains supplementary
information about the next-to-leading order process as well as about the
master integrals arising in the calculation.

\section{Generalities and Notation\label{sec:Notation}}
We study the Drell-Yan process of quark~($q$) anti-quark~($\bar{q}$) annihilation
into a charged lepton~($\ell$) pair
\begin{equation}
\label{eq:DY}
q(p_{1}) + \bar{q}(p_2) \rightarrow \ell^{-}(p_3) + \ell^{+}(p_4)
\end{equation}
where $p_1,p_2,p_3,p_4$ denote the momenta of the particles, which are all
considered as incoming with $p_1+p_2+p_3+p_4=0$. 
In the following we will use the Mandelstam variables
\begin{eqnarray}
\label{eq:mandel}
 s&=&(p_1+p_2)^2=(p_3+p_4)^2,\nonumber\\
 t&=&(p_1+p_3)^2=(p_2+p_4)^2,\nonumber\\
 u&=&(p_1+p_4)^2=(p_2+p_3)^2,
\end{eqnarray}
or $s_{ij} = (p_i + p_j)^2$
to express scalar products of the external momenta.

The differential cross section is given by
\begin{equation}
\label{eq:xsection}
{d\sigma^V\over d\Omega}={1\over64\*\pi^2\*s}{1\over 4\*\nc^2}
\sum_{\stackrel{\mbox{\tiny{spin}}}{\mbox{\tiny{color}}}}|\mathcal{M}|^2,
\end{equation}
where the symbol $\nc$ denotes the number of colors of SU($\nc$) and
$\mathcal{M}$ is the matrix element.  Within this work we consider only
virtual corrections to the Drell-Yan process. The perturbative expansion
of the corresponding squared matrix element is given by
\begin{equation}
\label{eq:decomp}
\sum_{\stackrel{\mbox{\tiny{spin}}}{\mbox{\tiny{color}}}}|\mathcal{M}|^2
                = \nc\*Q_q^2\*Q_\ell^2\*e^4\*\left( 
                                  A^{(0,0)}
                +\aepi          \*A^{(1,0)}
                +\aspi\*\CF     \*A^{(0,1)}
                +\aepi\aspi\*\CF\*A^{(1,1)}
                +\dots\right),
\end{equation}
where $Q_q$ and $Q_\ell$ are the electric charges of the initial state
quarks and the final state leptons in units of the elementary charge
$e$. The symbols $\alpha$ and $\alpha_s$ are the fine structure
constant and the strong coupling constant, respectively; $\CF =
(\nc^2-1)/(2\*\nc)$ denotes the Casimir operator of the fundamental
representation of SU($\nc$). The dots stand for higher order
corrections.  Here and in the following we will write the expansion of
any function of $\alpha$ and $\alpha_s$ as $f(\alpha,\alpha_s) =
\sum_{m,n} (\alpha /\pi)^m\,(\alpha_s / \pi)^n\,f^{(m,n)}$.  The
well-known leading order result $A^{(0,0)}$ of \eqn{eq:decomp} in
$d=4-2\*\ep$ space-time dimensions reads
\begin{equation}
\label{eq:LO}
A^{(0,0)}={8\over s^2}\*\left(t^2 + u^2 - s^2\*\ep\,\right).
\end{equation}
The one-loop
\qcd~\cite{Abad:1978nr,Altarelli:1978id,KubarAndre:1978uy,Harada:1979bj}
and one-loop \qed~corrections~\cite{Baur:1997wa} are known. For
completeness we will give the bare results for $A^{(1,0)}$ as well as
for $A^{(0,1)}$ in Appendix~\ref{sec:BareRes}, since they are needed for
the subtraction of the ultraviolet poles through the renormalization
procedure as well as for the identification of the infrared poles.

\section{Calculation of the bare process\label{sec:calculation}}
\begin{figure}[!ht]
\begin{minipage}{4cm}
\includegraphicsNEW{bb=62 541 275 689}{width=3.6cm}{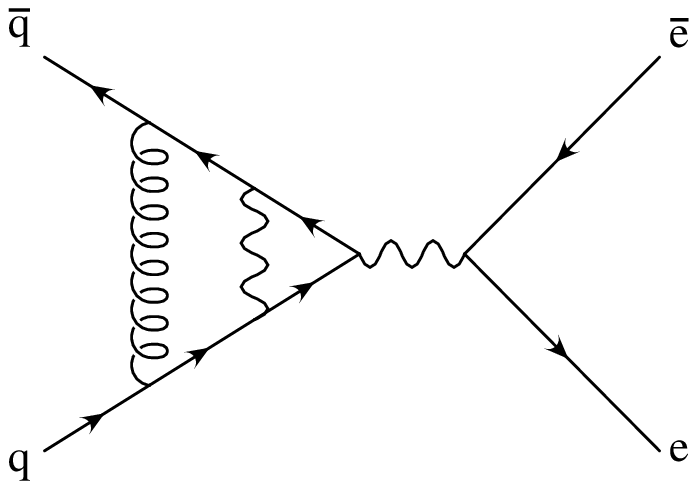}
\vspace*{-0.6cm}
\begin{center}
(a)
\end{center}
\end{minipage}
\hspace{0.4cm}
\begin{minipage}{4cm}
\includegraphicsNEW{bb=67 544 307 686}{width=4.0cm}{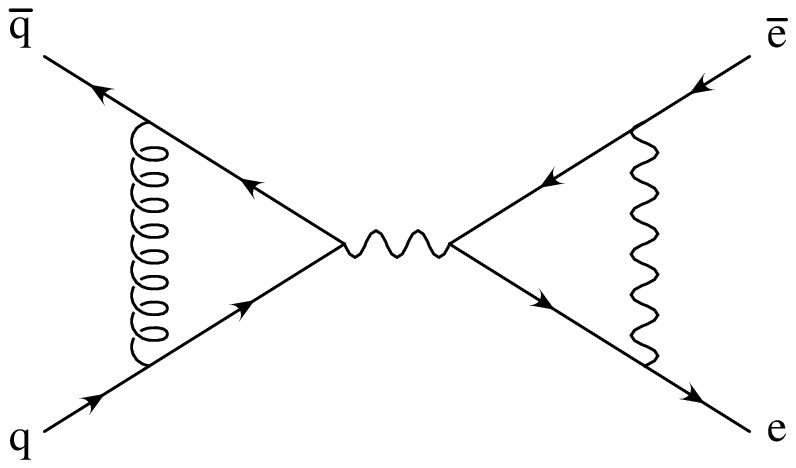}
\vspace*{-0.9cm}
\begin{center}
(b)
\end{center}
\end{minipage}
\hspace{0.4cm}
\begin{minipage}{4cm}
\includegraphicsNEW{bb=83 568 291 662}{width=4.3cm}{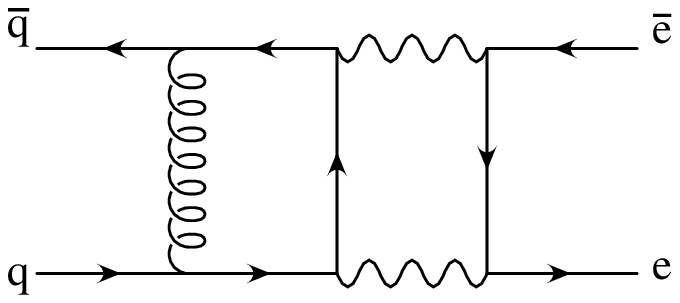}
\vspace*{-0.5cm}
\begin{center}
(c)
\end{center}
\end{minipage}\\[0.5cm]
\begin{minipage}{4cm}
\includegraphicsNEW{bb=52 541 318 689}{width=4.0cm}{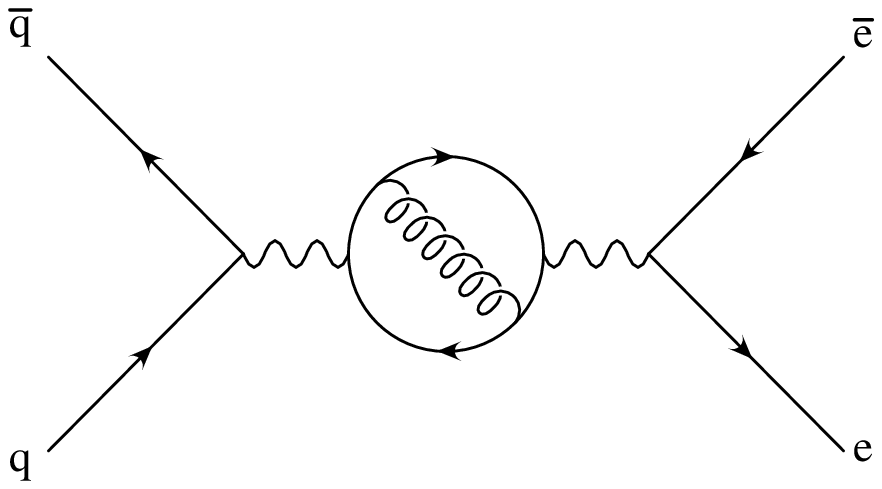}
\vspace*{-0.9cm}
\begin{center}
($\mbox{d}_1$)
\end{center}
\end{minipage}
\hspace{0.7cm}
\begin{minipage}{4cm}
\includegraphicsNEW{bb=59 544 320 686}{width=4.0cm}{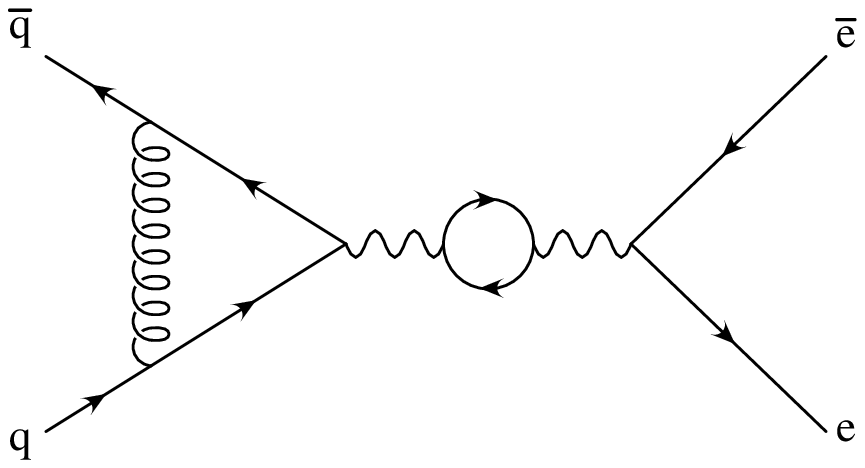}
\vspace*{-0.8cm}
\begin{center}
($\mbox{d}_2$)
\end{center}
\end{minipage}
\caption{Example diagrams which contribute to the two-loop mixed
  \qcd$\times$\qed\ corrections. The spiral lines denote gluons, the
  wavy lines are photons and the straight lines are fermions which can
  either be quarks in the initial state or leptons in the final
  state.\label{fig:samplediagrams}}
\end{figure}

Depending on the nature of the electromagnetic corrections, the
\qcd$\times$\qed\ corrections to Drell-Yan production can be 
broken up into four classes, which are: initial state corrections,
final state corrections, mixed initial and final state corrections,
and vacuum polarization corrections. Sample diagrams for each of these
four classes are shown in FIG~\ref{fig:samplediagrams}.

The initial state electromagnetic corrections consist of two-loop
corrections to the quark -- photon vertex.  A sample diagram of this
class of correction is shown in FIG.~\ref{fig:samplediagrams}(a). 
All of the diagrams that appear in this portion of the calculation are
topologically identical to diagrams that appear in the two-loop
\qcd\ corrections to Drell-Yan production.  By simultaneously
computing the two-loop \qcd\ corrections and verifying the known
result~\cite{Matsuura:1987wt}, we obtain a strong check on this part
of the calculation. 
We also include the interference of the one-loop \qcd\ correction
and the one-loop \qed\ initial state correction.  There is
but one diagram of each sort.

The final state virtual corrections are quite trivial, since the only
contributing two-loop diagram is the one shown in
FIG.~\ref{fig:samplediagrams}(b), 
which is just the product of two one-loop triangle diagrams.  The
virtual corrections to this channel also get a contribution from the
interference of the one-loop \qcd\ corrections and the one-loop final
state \qed\ corrections.  Again, there is but one diagram of each
sort.

The mixed initial and final state electromagnetic corrections 
are the most complicated terms in this calculation and the only
ones which involve the kinematic variables $t$ and $u$ in the loop
integrals.  A sample diagram is shown in
FIG.~\ref{fig:samplediagrams}(c). 
Even the  interference of the one-loop amplitudes is
relatively complicated as the electromagnetic part involves the sum of
two one-loop box integrals.

The vacuum polarization correction terms are also easy to compute as
the loops are simple two-loop propagator integrals, like that shown in
FIG.~\ref{fig:samplediagrams}($\mbox{d}_1$), or the product of a
one-loop propagator integral and a one-loop triangle as in
FIG.~\ref{fig:samplediagrams}($\mbox{d}_2$). 
The interference of one-loop amplitudes is again very simple as it
only involves a single vacuum polarization diagram interfered with the
single one-loop \qcd\ diagram.

We have performed two independent calculations of the virtual
corrections and find complete agreement. The Feynman diagrams are
generated with QGRAF~\cite{Nogueira:1991ex}.  The symbolic algebra
program FORM~\cite{Vermaseren:2000nd} is used to implement the Feynman
rules, to interfere the two-loop diagrams with the tree-level
contribution and to reduce the result to a set of Feynman integrals to
be determined. The calculation proceeds in two steps. In the first step
all the loop integrals are mapped onto a small set of master integrals
with the traditional integration-by-parts (IBP) method~\cite{Chetyrkin:1981qh}
in combination with Laporta's
algorithm~\cite{Laporta:1996mq,Laporta:2001dd}. In the second step these
master integrals are evaluated.

In one calculation, the integrals are reduced to master integrals
using the program REDUZE~\cite{Studerus:2009ye}.  In the second
calculation, the reduction has been performed with a
FORM~\cite{Vermaseren:2000nd,Vermaseren:2002rp,Tentyukov:2006ys} based
implementation which uses the packages Q2E and
EXP~\cite{Harlander:1997zb,Seidensticker:1999bb} to identify the
different topologies and to adopt the proper notation. The
program FERMAT~\cite{Fermat} is used to simplify the rational 
functions in the space time dimension $d$.

\begin{figure}[!ht]
\begin{minipage}[b]{5cm}
\includegraphicsNEW{bb=56 537 312 660}{width=5cm}{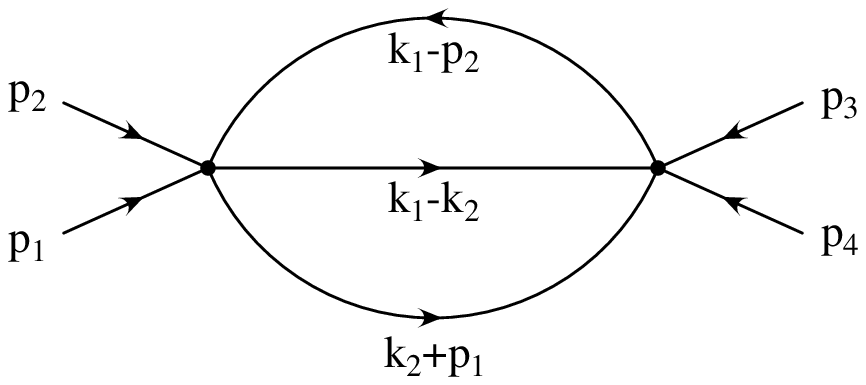}\\
\vspace*{-0.1cm}
\begin{center}
(a)
\end{center}
\end{minipage}
\hspace{0.3cm}
\begin{minipage}[b]{5cm}
\includegraphicsNEW{bb=56 550 312 649}{width=5cm}{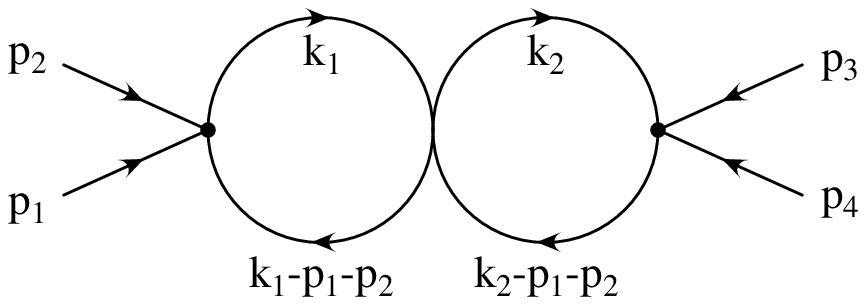}\\
\vspace*{-0.1cm}
\begin{center}
(b)
\end{center}
\end{minipage}
\hspace{0.3cm}
\begin{minipage}[b]{4cm}
\includegraphicsNEW{bb=0 0 205.02 229}{width=4.cm}{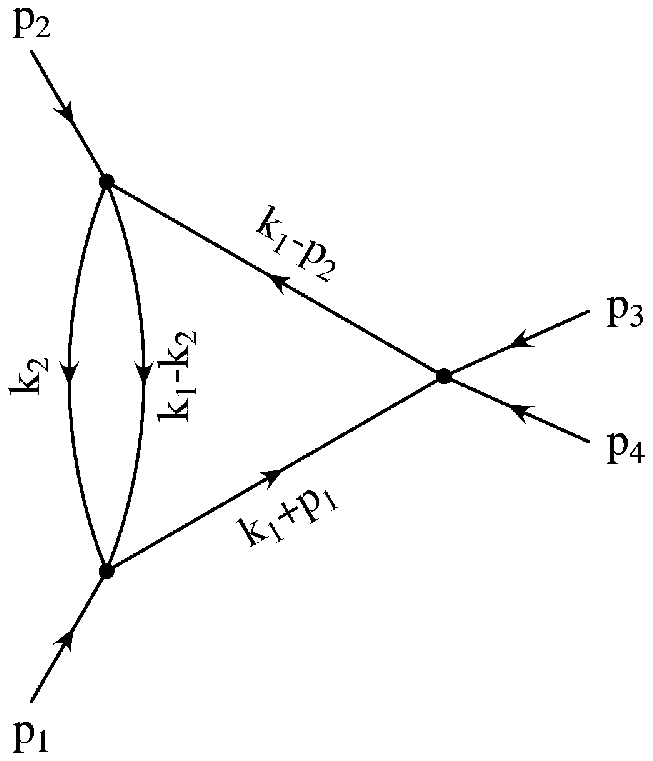}\\
\vspace*{-1.4cm}
\begin{center}
(c)
\end{center}
\end{minipage}\\[0.25cm]
\begin{minipage}{4cm}
\includegraphicsNEW{bb=90 518 281 709}{width=4.cm}{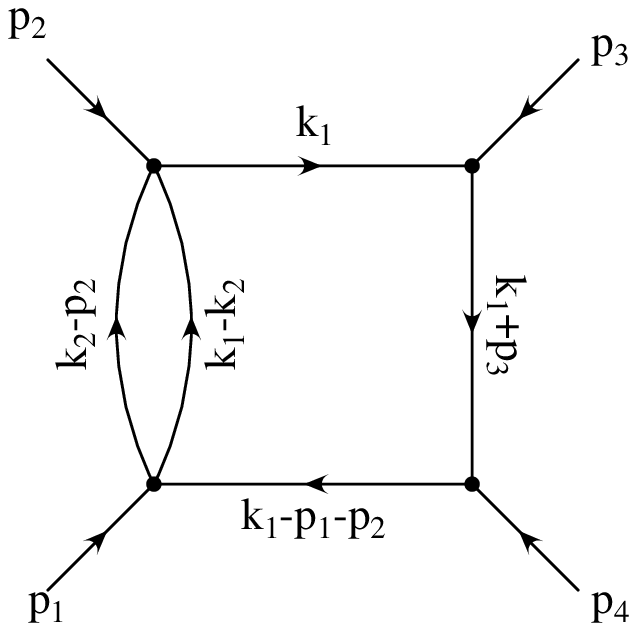}\\
\vspace*{-0.5cm}
\begin{center}
(d)
\end{center}
\end{minipage}
\hspace{0.8cm}
\begin{minipage}{4cm}
\includegraphicsNEW{bb=90 518 281 709}{width=4.cm}{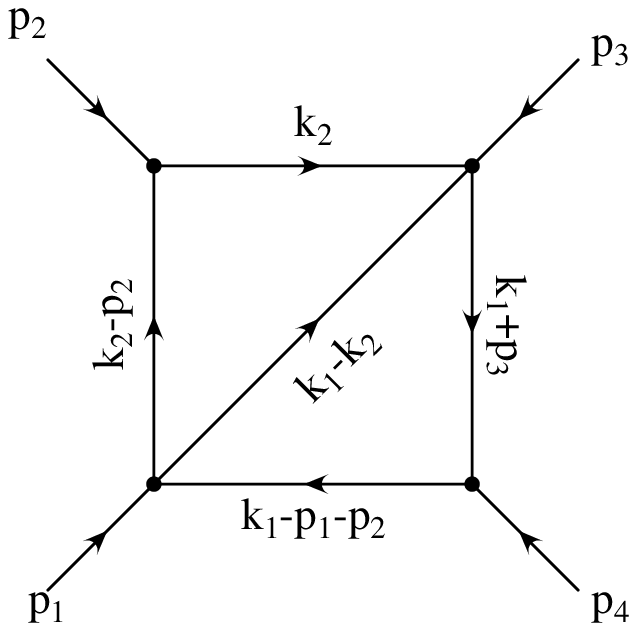}\\
\vspace*{-0.5cm}
\begin{center}
(e)
\end{center}
\end{minipage}
\hspace{0.8cm}
\begin{minipage}{3.2cm}
\includegraphicsNEW{bb=119 495 312 724}{width=3.2cm}{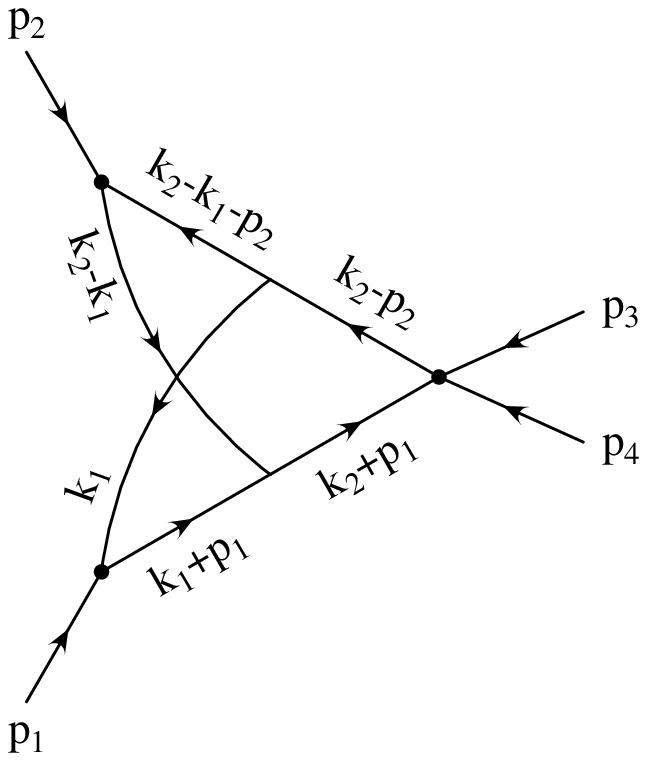}\\
\vspace*{-0.4cm}
\begin{center}
(f)
\end{center}
\end{minipage}\\[0.1cm]
\hspace{-3.5cm}
\begin{minipage}{2.35cm}
\includegraphicsNEW{bb=56 564 304 662}{height=2.35cm}{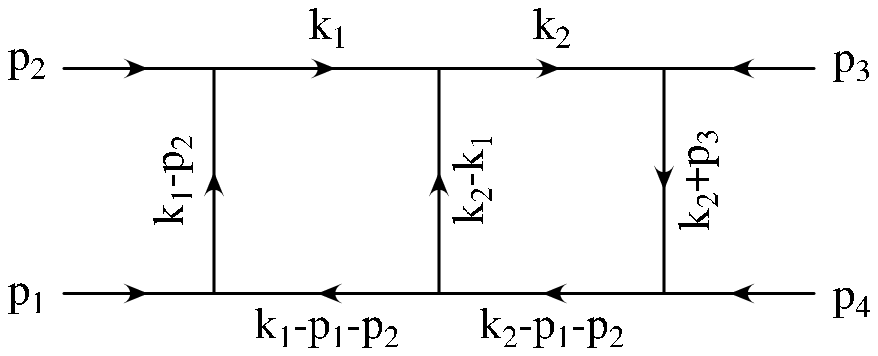}\\[0.25cm]
\hspace*{2.7cm}
($\mbox{g}_1$)
\end{minipage}
\hspace{5cm}
\raisebox{-0.05cm}{
\begin{minipage}{3.0cm}
{
\includegraphicsNEW{bb=56 564 304 662}{height=2.35cm}{DBox1}
}\\[0.04cm]
\begin{picture}(10,10)
\put(115,45){$\otimes\;(k_1+p_3)^2$}
\end{picture}\\[-0.25cm]
\hspace*{2.9cm}
($\mbox{g}_2$)
\end{minipage}}
\caption{The master topologies of the two-loop calculation ordered
  according to the number of internal lines. The arrow on the lines
  denotes the momentum flow. The symbols $k_1$ and $k_2$ are loop
  momenta. The integral representation of diagram $(g_2)$ has an
  additional irreducible scalar product $(k_1+p_3)^2$ in the numerator 
  indicated by the $\otimes$ symbol. \label{fig:MI}} 
\end{figure}

We find that at the two-loop level, all integrals can be expressed in
terms of eight master topologies which are shown in FIG.~\ref{fig:MI}.
All of the needed master integrals are known analytically in the
literature to sufficiently deep order in the $\ep$ expansion and will be
discussed in more detail in Appendix~\ref{sec:MI}.
At the end, the reductions and the evaluations of the master integrals
are substituted back into the FORM program to produce the final result.

\section{The pole structure of the process\label{sec:poles}}
\subsection{Ultraviolet Structure and Renormalization\label{sec:ultraviolet}}
We have performed renormalization in the \MSbar scheme.  Since we treat
all particles as being massless, we only need to renormalize the
couplings.  Coupling constant renormalization is governed by the
$\beta$-functions, 
\begin{eqnarray}
\label{eq:betafunctions1}
  \beta_\qed(\alpha,\alpha_s) = \mu^2\frac{d}{d\,\mu^2}\aepi &=& -
    \beta_\qed^{(2,0)}\aepi^2 - \beta_\qed^{(3,0)}\aepi^3 -
    \beta_\qed^{(2,1)}\aepi^2\,\aspi + \dots,\nonumber\\
  &&\beta_\qed^{(2,0)} = -\frac{1}{3}\Lx N_\ell\,Q_\ell^2 +
     \nc\,N_u\,Q_u^2 + \nc\,N_d\,Q_d^2\Rx\,,\nonumber\\
  &&\beta_\qed^{(3,0)} = -\frac{1}{4}\Lx N_\ell\,Q_\ell^4 +
     \nc\,N_u\,Q_u^4 + \nc\,N_d\,Q_d^4\Rx\,,\nonumber\\
  &&\beta_\qed^{(2,1)} = -\frac{1}{4}\CF\,\nc\Lx
     N_u\,Q_u^2 + N_d\,Q_d^2\Rx\,,\\
  \beta_\qcd(\alpha_s,\alpha) = \mu^2\frac{d}{d\,\mu^2}\aspi &=& -
    \beta_\qcd^{(0,2)}\aspi^2 - \beta_\qcd^{(0,3)}\aspi^3 -
    \beta_\qcd^{(1,2)}\aspi^2\,\aepi + \dots,\nonumber\\
  &&\beta_\qcd^{(0,2)} = \frac{11}{12}C_A - \frac{1}{3}T_f\,N_q,\nonumber\\
  &&\beta_\qcd^{(0,3)} = \frac{17}{24}C_A^2 - \frac{5}{12}C_A\,T_f\,N_q -
    \frac{1}{4}\CF\,T_f\,N_q,\nonumber\\
  &&\beta_\qcd^{(1,2)} = -\frac{1}{8}\Lx N_u\,Q_u^2 + N_d\,Q_d^2\Rx\,,
\label{eq:betafunctions2}
\end{eqnarray}
where $\mu$ is the renormalization scale, $N_u$ is the number of up-type
quarks, $N_d$ is the number of down-type quarks and $N_\ell$ is the
number of charged leptons, while $Q_u$, $Q_d$ and $Q_\ell$ are their
electric charges, $+\frac{2}{3}$, $-\frac{1}{3}$ and $-1$,
respectively. The symbol $C_A = \nc$ denotes the Casimir operator of the
adjoint representation of SU($\nc$) and $\displaystyle T_f
= 1/2$ is the normalization of the \qcd\ charge of the fundamental
representation. The bare and renormalized couplings are related by
\begin{eqnarray}
\aebarepi &=& \Lx\frac{e^{\gamma_E}}{4\,\pi}\Rx^{\ep}\,\aepi
    \LB1 - \aepi\frac{\beta^{(2,0)}}{\ep}
         - \aepi^2\Lx\frac{\beta^{(3,0)}}{2\,\ep}
         - \Lx\frac{\beta^{(2,0)}}{\ep}\Rx^2\Rx\right.\nonumber\\
       &&\hskip120pt\left.
      - \aepi\aspi\frac{\beta^{(2,1)}}{2\,\ep} + \dots\RB\,,\\
\asbarepi &=& \Lx\frac{e^{\gamma_E}}{4\,\pi}\Rx^{\ep}\,\aspi
    \LB1 - \aspi\frac{\beta^{(0,2)}}{\ep}
         - \aspi^2\Lx\frac{\beta^{(0,3)}}{2\,\ep}
         - \Lx\frac{\beta^{(0,2)}}{\ep}\Rx^2\Rx\right.\nonumber\\
       &&\hskip120pt\left.  - \aepi\aspi\frac{\beta^{(1,2)}}{2\,\ep} + \dots\RB\,,
\end{eqnarray}
where $e\simeq2.71828$ is Euler's number and $\gamma_E\simeq0.577216$ is the
Euler-Mascheroni constant.  Since the leading-order contribution to the
squared matrix element is of order $\alpha^2$ and we compute terms
through order $\alpha^3\,\alpha_s$, we need to keep
\qed\ renormalization terms proportional to $\beta^{(2,0)}$ and
$\beta^{(2,1)}$\cite{Gorishnii:1990vf}, while \qcd\ renormalization does
not contribute to our result.
The fine-structure constant can be converted, if needed, from the above \MSbar scheme
to the on-shell definition with a conversion factor. This conversion is
known to four-loop order in QED~\cite{Broadhurst:1991fi,Baikov:2008si}. 

\subsection{Infrared Structure\label{sec:infrared}}
An important check on our calculation is to verify that we have
obtained the correct infrared structure.  Some years ago,
Catani~\cite{Catani:1998bh} proposed a formula predicting the
leading poles ($\ep^{-4}$ through $\ep^{-2}$) of two-loop \qcd\
amplitudes.  At that time, the $\ep^{-1}$ poles were presumed to be
process dependent and therefore unpredictable.  Nonetheless, direct
calculations~\cite{Anastasiou:2000kg,Anastasiou:2000ue,
Anastasiou:2000mv,Bern:2002tk,Anastasiou:2002zn} showed that the
$\ep^{-1}$ terms seemed to follow a simple pattern based upon the
numbers of quarks and gluons that made up the external legs of the
amplitude.

Subsequently, Sterman and Tejeda-Yeomans~\cite{Sterman:2002qn}
reformulated Catani's observation and identified the origins of the
various terms.  They also identified the then-unknown term, the
second-order correction to the so-called ``soft anomalous dimension''
which prevented the prediction of the $\ep^{-1}$ terms.  Aybat, Dixon
and Sterman~\cite{Aybat:2006wq,Aybat:2006mz} have since computed the
two-loop corrections to the soft anomalous dimension, permitting the
prediction of the full infrared structure of two-loop \qcd\
amplitudes.

\subsection{The Infrared Structure of \qcd\ Amplitudes}
For a general $2\to\ n$ scattering process,
\begin{equation}
f_1(p_1,c_1) + f_2(p_2,c_2) \to\ f_3(p_3,c_3) + \dots + f_{n+2}(p_{n+2},c_{n+2})\,,
\end{equation}
where $f_i$ represent the flavors of the partons, $p_i$ their momenta
and $c_i$ their colors, we can write the amplitude as a vector in the
space of color tensors $\{(C_I)_{\{c_i\}}\}$
as~\cite{Catani:1996jh,Catani:1997vz,Catani:1998bh}
\begin{equation}
\ket{{\cal M}_{\bf f}\Lx p_i, \tfrac{Q^2}{\mu^2}, \alpha_s(\mu^2),
  \ep\Rx} \equiv \sum_{L} {\cal M}_{{\bf f},L}\Lx p_i,
  \tfrac{Q^2}{\mu^2}, \alpha_s(\mu^2), \ep\Rx\times (C_L)_{\{c_i\}}\,,
\end{equation}
where $Q$ is an (arbitrary) overall scale and $\mu$ is the
renormalization scale.

In the formulation of
Refs.~\cite{Sterman:2002qn,Aybat:2006wq,Aybat:2006mz}, a renormalized
amplitude may be factorized into three functions: the jet function
${\cal J}_{\bf f}$, which describes the collinear dynamics of the
external partons that participate in the collision; the soft function
${\bf S_f}$, which describes soft exchanges between the external
partons; and the hard-scattering function $\ket{H_{\bf f}}$, which
describes the short-distance scattering process
\begin{equation}
\ket{{\cal M}_{\bf f}\Lx p_i,\tfrac{Q^2}{\mu^2},\alpha_s(\mu^2),\ep\Rx} =
   {\cal J_{\bf f}}\Lx\alpha_s(\mu^2),\ep\Rx\
    {\bf S_{f}}\Lx p_i,\tfrac{Q^2}{\mu^2},\alpha_s(\mu^2),\ep\Rx\
    \ket{H_{\bf f}\Lx p_i,\tfrac{Q^2}{\mu^2},\alpha_s(\mu^2)\Rx}\,.
\label{eqn:qcdfact}
\end{equation}
The notation indicates that $\ket{H_{\bf f}}$ is a vector and ${\bf
  S_f}$ is a matrix in color space. As with any factorization, there
is considerable freedom to move terms about from one function to the
others.  It is convenient~\cite{Aybat:2006wq,Aybat:2006mz} to define
the jet and soft functions, ${\cal J}_{\bf f}$ and ${\bf S_f}$, so
that they contain all of the infrared poles but only contain infrared
poles, while all infrared finite terms are absorbed into $\ket{H_{\bf
    f}}$.

\subsubsection{The Jet Function}
The jet function ${\cal J}_{\bf f}$ is found to be the product of
individual jet functions ${\cal J}_{f_i}$ for each of the external
partons,
\begin{equation}
{\cal J}_{\bf f}\Lx\alpha_s(\mu^2),\ep\Rx = \prod_{i\in{\bf{f}}}\
   {\cal J}_{i}\Lx\alpha_s(\mu^2),\ep\Rx\,.
\end{equation}
Each individual jet function is naturally defined in terms of the
Sudakov form factor~\cite{Sterman:2002qn},
\begin{equation}
{\cal J}_{i}\Lx\alpha_s(\mu^2),\ep\Rx
    = {\cal J}_{\bar{\imath}}\Lx\alpha_s(\mu^2),\ep\Rx
    \sim \LB{\cal M}^{[i\,\bar\imath\to1]}\Lx\alpha_s(\mu^2),\ep\Rx\RB^{1/2}
\end{equation}
The all-orders expression for the square root of the Sudakov form factor
is~\cite{Collins:1989bt,Magnea:1990zb,Magnea:2000ss,Magnea:2000ep}
\begin{equation}
\begin{split}
J_{i}\Lx\alpha_s(\mu^2),\ep\Rx &= \exp\left\{\frac{1}{4}\int_0^{\mu^2}
  \frac{d\,\xi^2}{\xi^2}\bigg[{\cal K}_i\Lx\alpha_s(\mu^2),\ep\Rx
   + {\cal G}_i\Lx-1,\bar{\alpha}_s\Lx\tfrac{\mu^2}{\xi^2},
  \alpha_s(\mu^2),\ep\Rx,\ep\Rx\right.\\
 & \left.\hskip100pt+ \frac{1}{2}\int_{\xi^2}^{\mu^2}
  \frac{d\,\tilde\mu^2}{\tilde\mu^2}\gamma_{K\,i}\Lx\bar\alpha_s\Lx
  \tfrac{\mu^2}{\tilde\mu^2},\alpha_s(\mu^2),\ep\Rx\Rx\bigg]\right\}\,.
\end{split}
\end{equation}
The functions ${\cal K}_i$, ${\cal G}_i$ and $\gamma_{K\,i}$ are
anomalous dimensions that can be determined from fixed-order
calculations of the Sudakov form factors for quarks and
gluons~\cite{Gonsalves:1983nq,
Kramer:1986sg,Matsuura:1987wt,Matsuura:1989sm,Harlander:2000mg,
Moch:2005id,Moch:2005tm}.  Note that $\gamma_{K\,i}$ is the cusp
anomalous dimension and ${\cal K}_i$ is determined, order by order,
from $\gamma_{K\,i}$.  While the ${\cal K}_i$ are pure pole terms, the
${\cal G}_i$ contain terms at higher order in $\ep$.

The jet functions ${\cal J}_{f_i}$ keep only the infrared poles from
the logarithm of the form factor. The expansion of the jet function to
second order in $\alpha_s$ is
\begin{equation}
\begin{split}
\ln{\cal J}_i\Lx\alpha_s(\mu^2),\ep\Rx &=
  -\aspi\LB\frac{1}{8\,\ep^2}\gamma_{K\,i}^{(0,1)} +
  \frac{1}{4\,\ep}{\cal G}_i^{(0,1)}(\ep)\RB\\
  &\quad + \aspi^2\left\{
   \frac{\beta_\qcd^{(0,2)}}{8}\frac{1}{\ep^2}\LB\frac{3}{4\,\ep}\gamma_{K\,i}^{(0,1)}
    + {\cal G}_i^{(0,1)}(\ep)\RB -
    \frac{1}{8}\LB\frac{\gamma_{K\,i}^{(0,2)}}{4\,\ep^2} + \frac{{\cal
	G}_i^{(0,2)}(\ep)}{\ep}\RB  \right\} + \dots
\end{split}
\end{equation}
where
\begin{equation}
\begin{split}
    \gamma_{K\,i}^{(0,1)} &= 2\,C_i,\quad
    \gamma_{K\,i}^{(0,2)} = C_i\,K = C_i\LB C_A\Lx\frac{67}{18} -
    \zeta_2\Rx - \frac{10}{9}T_f\,N_q\RB,\quad C_q \equiv \CF,\quad
    C_g \equiv C_A,\\
  {\cal G}_{q}^{(0,1)} &= \frac{3}{2}\CF + \frac{\ep}{2}\CF\Lx8-\zeta_2\Rx, 
  \qquad {\cal G}_{g}^{(0,1)} = 2\,\beta_\qcd^{(0,2)} - \frac{\ep}{2}C_A\,\zeta_2,\\
  {\cal G}_{q}^{(0,2)} &= \CF^2\Lx\frac{3}{16} - \frac{3}{2}\zeta_2 +
    3\,\zeta_3\Rx + \CF\,C_A\Lx\frac{2545}{432} + \frac{11}{12}\zeta_2
    - \frac{13}{4}\zeta_3\Rx - \CF\,T_f\,N_q\Lx\frac{209}{108} +
    \frac{1}{3}\zeta_2\Rx,\\
  {\cal G}_{g}^{(0,2)} &= 4\,\beta_\qcd^{(0,3)} + C_A^2\Lx\frac{10}{27} -
    \frac{11}{12}\zeta_2 - \frac{1}{4}\zeta_3\Rx +
    C_A\,T_f\,N_q\Lx\frac{13}{27} + \frac{1}{3}\zeta_2\Rx +
    \frac{1}{2}\CF\,T_f\,N_q\,,
\label{eqn:qcdconsts}
\end{split}
\end{equation}
$N_q$ is the number of quark flavors and
$\z{n}=\sum_{k=1}^{\infty}1/k^n$ represents the Riemann zeta-function
of integer argument $n$. The coefficients of the $\beta$-functions are
given in \eqns{eq:betafunctions1}{eq:betafunctions2}. Even though the
${\cal G}_i$ have terms at higher order in $\ep$, we only keep terms
in the expansion that contribute poles to $\ln{\cal J}_i$.

\subsubsection{The Soft Function}
Like the jet function, the soft function can be defined in terms of
eikonal amplitudes and is determined entirely by the soft anomalous
dimension matrix ${\bm\Gamma}_{S_f}$,
\begin{equation}
\begin{split}
{\bf S_f}\Lx p_i,\tfrac{Q^2}{\mu^2},\alpha_s(\mu^2),\ep\Rx &= 
   {\rm P}\,\exp\left\{-\frac{1}{2}\int_0^{\mu^2}
   \frac{d\,\bar\mu^2}{\bar\mu^2}{\bm\Gamma}_{S_f}\Lx\tfrac{s_{ij}}{\mu^2},
   \bar\alpha_s\Lx\tfrac{\mu^2}{\tilde\mu^2},\alpha_s(\mu^2),\ep\Rx\Rx\right\}\\
  &= 1 + \frac{1}{2\,\ep}\aspi{\bm\Gamma}_{S_f}^{(0,1)} +
   \frac{1}{8\,\ep^2}\aspi^2{\bm\Gamma}_{S_f}^{(0,1)}\times{\bm\Gamma}_{S_f}^{(0,1)}\\
  &\qquad- \frac{\beta_{\qcd}^{(0,2)}}{4\,\ep^2}\aspi^2{\bm\Gamma}_{S_f}^{(0,1)}
   + \frac{1}{4\,\ep}\aspi^2{\bm\Gamma}_{S_f}^{(0,2)}\,.
\end{split}
\end{equation}
In the color-space notation of
Refs.~\cite{Catani:1996jh,Catani:1997vz,Catani:1998bh}, the soft
anomalous dimension is given by~\cite{Aybat:2006wq,Aybat:2006mz}
\begin{equation}
{\bm\Gamma}_{S_f}^{(0,1)} = \frac{1}{2}\,\sum_{i\in{\bf f}}\ \sum_{j\ne i}
   {\bf T}_i\cdot{\bf T}_j\,\ln\Lx\frac{\mu^2}{-s_{ij}}\Rx,\qquad
 {\bm\Gamma}_{S_f}^{(0,2)} = \frac{K}{2}{\bm\Gamma}_{S_f}^{(0,1)}\,,
\label{eqn:qcdsoftanomdim}
\end{equation}
where $K = C_A\Lx67/18-\zeta_2\Rx - 10\,T_f\,N_q/9$ is the same constant
that relates the one- and two-loop cusp anomalous dimensions. The ${\bf
  T}_i$ are the color generators in the representation of parton $i$,
multiplied by $\pm1$, depending on the whether the parton is a particle
or antiparticle and whether it is incoming or outgoing.  In particular,
outgoing quarks and gluons and incoming anti-quarks are multiplied by
$+1$, while incoming quarks and gluons and outgoing anti-quarks are
multiplied by $-1$.  The conservation of color-charge is enforced by the
identity $\sum_i\,{\bf T}_i = 0$.  Another useful identity is that ${\bf
  T}_i\cdot{\bf T}_i = C_i$.

\subsection{The Infrared Structure of \qed\ Amplitudes}
It has been found that the same factorization described in
\eqn{eqn:qcdfact} can be applied to pure \qed\
amplitudes~\cite{Bern:2000ie,Anastasiou:2002zn}.  The two-loop
amplitudes for Bhabha scattering and for $e^+\,e^-\to\gamma\gamma$ in
massless \qed\ were found to obey the factorization formula of
Catani~\cite{Catani:1998bh} once the proper adjustments are made to
transform the \qcd\ anomalous dimensions into \qed\ anomalous
dimensions.

The changes are as follows.  The factors of the adjoint representation
Casimir, $C_A$, originate from the gluon self interactions.  As
photons have no self interactions, $C_A$ is set to zero.  The
fundamental representation Casimir, $\CF$ is replaced by the squared
electric charge of the fermion, $\CF \to Q_i^2$.  The factors of
$T_f\,N_q$ originate from inserting fermion bubbles into the gluon
propagators.  In \qed, the different types of fermions would be
weighted by the squares of their electric charges, $T_f\,N_q \to
\nc\,N_u\,Q_u^2 + \nc\,N_d\,Q_d^2 + N_\ell\,Q_\ell^2$. In the soft
anomalous dimension, the color charge matrices ${\bf T}_i$ are
replaced by the (scalar) electric charges $Q_i$.  With these changes,
the anomalous dimensions for the \qed\ jet function are
\begin{equation}
\begin{split}
    \gamma_{K\,i}^{(1,0)} &= 2\,Q_i^2,\qquad
    \gamma_{K\,i}^{(2,0)} = Q_i^2\,K^\qed =
    \frac{10}{3}\,Q_i^2\,\beta_\qed^{(2,0)},\\
  {\cal G}_{f}^{(1,0)} &= \frac{3}{2}Q_f^2 + \frac{\ep}{2}Q_f^2\Lx8-\zeta_2\Rx,
  \qquad 
  {\cal G}_{f}^{(2,0)} = Q_f^4\Lx\frac{3}{16} - \frac{3}{2}\zeta_2 +
    3\,\zeta_3\Rx + Q_f^2\,\beta_\qed^{(2,0)}\Lx\frac{209}{36} +
    \zeta_2\Rx,\\
  {\cal G}_{\gamma}^{(1,0)} &= 2\,\beta_\qed^{(2,0)},\qquad{\cal
    G}_{\gamma}^{(2,0)} = 2\,\beta_\qed^{(3,0)}\,,
\label{eqn:qedconsts}
\end{split}
\end{equation}
while the \qed\ contribution to the soft anomalous dimension is
\begin{equation}
{\bm\Gamma}_{S_f}^{(1,0)} = \frac{1}{2}\,\sum_{i\in{\bf f}}\ \sum_{j\ne i}
   Q_i\,Q_j\,\ln\Lx\frac{\mu^2}{-s_{ij}}\Rx,\qquad
 {\bm\Gamma}_{S_f}^{(2,0)} = \frac{K^\qed}{2}{\bm\Gamma}_{S_f}^{(1,0)}
  = \frac{5}{3}\beta_\qed^{(2,0)}\,{\bm\Gamma}_{S_f}^{(1,0)}\,.
\label{eqn:qedsoftanomdim}
\end{equation}

Using these parameters, one can predict the infrared structure of
two-loop \qed\ amplitudes, where the analog rules as in
\eqn{eqn:qcdsoftanomdim} apply for the signs. When comparing to the results of
Refs.~\cite{Bern:2000ie,Anastasiou:2002zn}, one must account for the
fact that those calculations are in the context of pure \qed,
involving only leptons and photons.  As the universality of the
$\ep^{-1}$ terms had not yet been established, the (color
diagonal) $H^{(2)}$ factors for electrons and photons was quoted as
\begin{equation*}
\begin{split}
H_e^{(2)} &= - \Lx\frac{3}{8} - 3\,\zeta_2 + 6\,\zeta_3\Rx +
N_f^\prime\Lx-\frac{25}{54} + \frac{1}{2}\zeta_2\Rx,\\
H_\gamma^{(2)} &= \frac{20}{27}N_f^{\prime\,2} + N_f^\prime,\hskip
130pt{}
\end{split}
\end{equation*}
where $N_f^\prime \equiv N_\ell\,Q_\ell^2$.  Transforming the results
above into the notation of Ref.~\cite{Catani:1998bh}, we find that the
$H^{(2)}$ terms may be more generally written as
\begin{equation}
\begin{split}
H_f^{(2)} &= - Q_f^4\,\Lx\frac{3}{8} - 3\,\zeta_2 + 6\,\zeta_3\Rx +
   Q_f^2\,\beta_\qed^{(2,0)}\Lx\frac{25}{18} + \frac{3}{2}\zeta_2\Rx,\\
H_\gamma^{(2)} &= \frac{20}{3}\Lx\beta_\qed^{(2,0)}\Rx^2 -
   4\,\beta_\qed^{(3,0)}\,,
\end{split}
\end{equation}
where the subscript $f$ indicates any charged fermion -- lepton or
quark.  With these modifications, we find complete agreement with the
results of Refs.~\cite{Bern:2000ie,Anastasiou:2002zn}.

\subsection{The Infrared Structure of \qcd $\times$ \qed\ Amplitudes\label{sec:IRQCDQED}}
\begin{figure}[ht]
\includegraphicsNEW{bb=0 0 188 66}{width=6.cm}{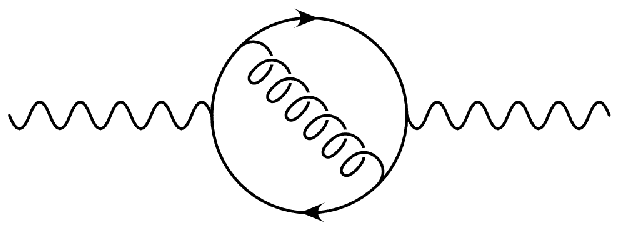}\hskip50pt
\includegraphicsNEW{bb=0 0 188 66}{width=6.cm}{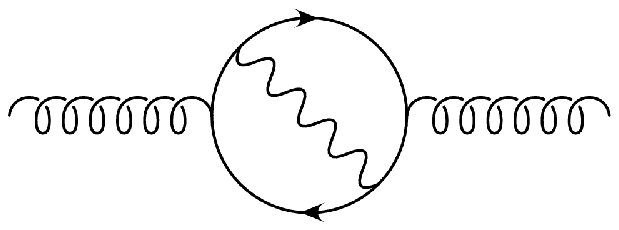}
\caption{Mixed second order contributions to the \qed\ and \qcd\ $\beta$-functions.
\label{fig:mixedbubbles}}
\end{figure}

The leading terms in the infrared structure of \qcd\ $\times$ \qed\
corrections will come from the overlap of the one-loop terms for pure
\qcd\ and pure \qed.  The intrinsically \qcd$\times$ \qed\ terms
will be second-order contributions to the jet and soft functions.
Based upon the way the parameters were determined for \qed, we can
make conjectures about the parameters for \qcd$\times$ \qed.  Since
the generators for photons and gluons commute, we should again set the
$C_A$ terms to zero.  We need to be a little more careful about the
$N_f$ terms, however.  Our approach is to tie the $N_f$ terms to the
coefficients of the $\beta$-functions.  The reason for this is that
when the $N_f$ term is part of the leading term in a $\beta$-function,
it represents the insertion of a fermion bubble into a gauge boson
propagator.  Because the charge matrix of \qcd\ is traceless, the
bubble cannot connect a photon to a gluon and therefore these terms
cannot contribute to a second-order mixed correction.  When the $N_f$
term is part of a second-order term in a $\beta$-function, however, it
represents a term like those shown in FIG.~\ref{fig:mixedbubbles},
which can represent a second-order mixed correction.
Examining the two-loop anomalous dimensions in
Eqs.~(\ref{eqn:qcdconsts}) and~(\ref{eqn:qcdsoftanomdim}), we see that
the second order corrections to the cusp and soft anomalous dimensions
are proportional to $K = C_A\Lx67/18-\zeta_2\Rx - 10\,T_f\,N_q/9 =
\Lx2/3-\zeta_2\Rx\,C_A + 10/3\,\beta_\qcd^{(0,2)}$.  Since we have
argued that neither non-Abelian nor first-order $\beta$-function
corrections can contribute to second-order mixed corrections, we
conclude that there are no mixed corrections to the cusp and soft
anomalous dimensions at this order.  That leaves only the ${\cal
  G}_{i}$ terms.  By the same reasoning as for the cusp and soft
anomalous dimensions, we set the $C_A$ and $N_q$ terms to zero in
forming ${\cal G}_q^{(1,1)}$, but we predict that the $\CF^2$ term
should be transformed into $\CF\,Q_q^2$.  For ${\cal
  G}_{g\,,\gamma}^{(1,1)}$, we again drop the non-Abelian and
first-order $\beta$-functions, but we predict that we should keep the
second-order $\beta$-function terms to obtain ${\cal G}_{g}^{(1,1)} =
2\beta_\qcd^{(1,2)}$ and ${\cal G}_{\gamma}^{(1,1)} =
2\beta_\qed^{(2,1)}$.

We can thus write combined expressions for the jet and soft functions
which we claim are valid through second order in both \qcd\ and \qed,
\begin{eqnarray}
\ln{\cal J}_i(\alpha(\mu^2),\alpha_s(\mu^2),\ep) &\!\!=\!\!&
       - \aepi\LB\frac{1}{8\,\ep^2}\gamma_{K\,,i}^{(1,0)} +
     \frac{1}{4\,\ep}{\cal G}_i^{(1,0)}(\ep)\RB
       - \aspi\LB\frac{1}{8\,\ep^2}\gamma_{K\,,i}^{(0,1)} +
     \frac{1}{4\,\ep}{\cal G}_i^{(0,1)}(\ep)\RB\nonumber\\
    && + \aepi^2\left\{\frac{\beta_\qed^{(2,0)}}{8\,\ep^2}\LB
       \frac{3}{4\,\ep}\gamma_{K\,i}^{(1,0)}
              + {\cal G}_i^{(1,0)}(\ep)\RB
       - \frac{1}{8}\LB\frac{1}{4\,\ep^2}\gamma_{K\,i}^{(2,0)} +
     \frac{1}{\ep}{\cal G}_i^{(2,0)}\RB\right\}\nonumber\\
    && + \aspi^2\left\{\frac{\beta_\qcd^{(0,2)}}{8\,\ep^2}\LB
       \frac{3}{4\,\ep}\gamma_{K\,i}^{(0,1)}
              + {\cal G}_i^{(0,1)}(\ep)\RB
       - \frac{1}{8}\LB\frac{1}{4\,\ep^2}\gamma_{K\,i}^{(0,2)} +
     \frac{1}{\ep}{\cal G}_i^{(0,2)}\RB\right\}\nonumber\\
    && - \aepi\aspi\frac{1}{4\ep}{\cal G}_i^{(1,1)} + \dots\,,
\end{eqnarray}
and
\begin{eqnarray}
{\bf S_f}
\!\!\!\!&\!\!\!\!\!\!&\!\!\!\!
\Lx p_i,\tfrac{Q^2}{\mu^2},\alpha(\mu^2),\alpha_s(\mu^2),\ep\Rx =
   1 + \frac{1}{2\,\ep}\aepi{\bm\Gamma}_{S_f}^{(1,0)}
     + \frac{1}{2\,\ep}\aspi{\bm\Gamma}_{S_f}^{(0,1)}\nonumber\\
  &&+ \frac{1}{8\,\ep^2}\aepi^2{\bm\Gamma}_{S_f}^{(1,0)}
       \times{\bm\Gamma}_{S_f}^{(1,0)}
    + \frac{1}{8\,\ep^2}\aspi^2{\bm\Gamma}_{S_f}^{(0,1)}
       \times{\bm\Gamma}_{S_f}^{(0,1)}
    + \frac{1}{4\,\ep^2}\aepi\aspi{\bm\Gamma}_{S_f}^{(1,0)}
       \times{\bm\Gamma}_{S_f}^{(0,1)}\nonumber\\
  && - \frac{\beta_\qed^{(1,0)}}{4\,\ep^2}\aepi^2{\bm\Gamma}_{S_f}^{(1,0)}
     - \frac{\beta_\qcd^{(0,2)}}{4\,\ep^2}\aspi^2{\bm\Gamma}_{S_f}^{(0,1)}
    + \frac{1}{4\,\ep}\aepi^2{\bm\Gamma}_{S_f}^{(2,0)}
    + \frac{1}{4\,\ep}\aspi^2{\bm\Gamma}_{S_f}^{(0,2)}\,,
\end{eqnarray}
with
\begin{align}
\gamma_{K\,i}^{(1,0)} &= 2\,Q_i^2\,,\qquad \gamma_{K\,i}^{(2,0)} =
   \gamma_{K\,i}^{(1,0)}\,K_\qed\ = \frac{10}{3}Q_i^2\,\beta_\qed^{(2,0)}\,,\nonumber\\
\gamma_{K\,i}^{(0,1)} &= 2\,C_i\,,\qquad \gamma_{K\,i}^{(0,2)} =
   \gamma_{K\,i}^{(0,1)}\,K_\qcd\ = \LB\Lx\frac{2}{3} - \zeta_2\Rx\,C_A
   + \frac{10}{3}\,\beta_\qcd^{(2,0)}\RB\,C_i\,,\nonumber\\
{\cal G}_f^{(1,0)} &= \frac{3}{2}Q_f^2 + \frac{\ep}{2}Q_f^2\Lx8 -
   \zeta_2\Rx\,,\qquad {\cal G}_f^{(2,0)} = Q_f^4\Lx\frac{3}{16} -
   \frac{3}{2}\zeta_2 + 3\,\zeta_3\Rx +
   Q_f^2\,\beta_\qed^{(2,0)}\,\Lx\frac{209}{36} + \zeta_2\Rx\,,\nonumber\\
{\cal G}_\gamma^{(1,0)} &= 2\,\beta_\qed^{(2,0)}\,,\hskip83pt
   {\cal G}_\gamma^{(2,0)} = 2\,\beta_\qed^{(3,0)}\,,\nonumber\\
{\cal G}_f^{(0,1)} &= \frac{3}{2}\CF + \frac{\ep}{2}\CF\Lx8 -
   \zeta_2\Rx\,,\nonumber\\
{\cal G}_f^{(0,2)} &= \CF^2\Lx\frac{3}{16} -
   \frac{3}{2}\zeta_2 + 3\,\zeta_3\Rx + \CF\,C_A\Lx\frac{41}{72} -
   \frac{13}{4}\zeta_3\Rx  + \beta_\qcd^{(0,2)}\Lx\frac{209}{36} + \zeta_2\Rx\,,\nonumber\\
{\cal G}_g^{(0,1)} &= 2\,\beta_\qcd^{(0,2)} -
   \frac{\ep}{2}C_A\,\zeta_2\,,\qquad
   {\cal G}_g^{(0,2)} = 2\,\beta_\qcd^{(0,3)} + \Lx\frac{19}{18} -
   \zeta_2\Rx C_A\,\beta_\qcd^{(0,2)} +
   \Lx\frac{59}{72}-\frac{1}{4}\zeta_3\Rx C_A^2\,,\nonumber\\
{\cal G}_f^{(1,1)} &= \CF\,Q_f^2\Lx\frac{3}{16} -\frac{3}{2}\zeta_2 +
   3\,\zeta_3\Rx\,,\qquad {\cal G}_\gamma^{(1,1)} =
   2\,\beta_\qed^{(2,1)}\,,\qquad
   {\cal G}_g^{(1,1)} = 2\,\beta_\qcd^{(1,2)}\,,\nonumber\\
{\bm\Gamma}_{S_f}^{(1,0)} &= \frac{1}{2}\,\sum_{i\in{\bf f}}\ \sum_{j\ne i}
   Q_i\,Q_j\,\ln\Lx\frac{\mu^2}{-s_{ij}}\Rx,\qquad
 {\bm\Gamma}_{S_f}^{(2,0)} = \frac{K^\qed}{2}{\bm\Gamma}_{S_f}^{(1,0)}
  = \frac{5}{3}\beta_\qed^{(2,0)}\,{\bm\Gamma}_{S_f}^{(1,0)}\,,\nonumber\\
{\bm\Gamma}_{S_f}^{(0,1)} &= \frac{1}{2}\,\sum_{i\in{\bf f}}\ \sum_{j\ne i}
   {\bf T}_i\cdot{\bf T}_j\,\ln\Lx\frac{\mu^2}{-s_{ij}}\Rx\,,\quad
 {\bm\Gamma}_{S_f}^{(0,2)} = \frac{K^\qcd}{2}{\bm\Gamma}_{S_f}^{(0,1)}
  = \LB\Lx\frac{1}{3} - \frac{1}{2}\zeta_2\Rx\,C_A
   + \frac{5}{3}\,\beta_\qcd^{(2,0)}\RB\,{\bm\Gamma}_{S_f}^{(0,1)}\,.
\end{align}

\subsection{The Infrared Structure of the Drell-Yan Amplitude\label{sec:IRstructure}}
We can now examine our result for the Drell-Yan amplitude to see if we
match the expected infrared structure.  We start from the
factorization formula, \eqn{eqn:qcdfact}, and expand both sides in
powers of $\alpha$ and $\alpha_s$,
\begin{eqnarray}
\label{eq:IRpoles1}
  \ket{{\cal M}_{DY}} 
    &=& {\cal J}_{DY}\,{\bf S}_{DY}\,\ket{H_{DY}}\\
    &=&\ket{{\cal M}^{(1,0)}_{DY}} + \aepi\ket{{\cal M}^{(2,0)}_{DY}} 
     + \aspi\ket{{\cal M}^{(1,1)}_{DY}} + \aepi\aspi\ket{{\cal
       M}^{(2,1)}_{DY}}\nonumber\\
    &=& \ket{H^{(1,0)}_{DY}}
     + \aepi\Lx{\cal J}^{(1,0)}_{DY}\,\ket{H^{(1,0)}_{DY}}
     + {\bf S}^{(1,0)}_{DY}\,\ket{H^{(1,0)}_{DY}}
     + \ket{H^{(2,0)}_{DY}}\Rx\nonumber\\
    &+& \aspi\Lx{\cal J}^{(0,1)}_{DY}\,\ket{H^{(1,0)}_{DY}}
     + {\bf S}^{(0,1)}_{DY}\,\ket{H^{(1,0)}_{DY}}
     + \ket{H^{(1,1)}_{DY}}\Rx\nonumber\\
    &+&\aepi\aspi\LB\Lx{\cal J}^{(1,1)}_{DY}
     + {\cal J}^{(1,0)}_{DY}\,{\bf S}^{(0,1)}_{DY}
     + {\cal J}^{(0,1)}_{DY}\,{\bf S}^{(1,0)}_{DY}
     + {\bf S}^{(1,1)}_{DY}\Rx
          \ket{H^{(1,0)}_{DY}}\right.\nonumber\\
    &&\quad\left.
     + \Lx{\cal J}^{(1,0)}_{DY}
     + {\bf S}^{(1,0)}_{DY}\Rx\,\ket{H^{(1,1)}_{DY}}
     + \Lx{\cal J}^{(0,1)}_{DY}
     + {\bf S}^{(0,1)}_{DY}\Rx\,\ket{H^{(2,0)}_{DY}}
     + \ket{H^{(2,1)}_{DY}}\RB\,.
\label{eq:IRpoles}
\end{eqnarray}
Because of the trivial color structure of the Drell-Yan amplitude, the
soft anomalous dimension matrix is proportional to the unit matrix and
may be treated as a scalar function. The squared matrix element of
\eqn{eq:decomp} is related to the decomposition of the amplitude in
\eqn{eq:IRpoles1} by
$\sum_{\stackrel{\mbox{\tiny{spin}}}{\mbox{\tiny{color}}}}|\mathcal{M}|^2=
\Lx{e^{\gamma_E}/(4\,\pi)}\Rx^{2\*\ep}
\aepi^2\bra{{\cal M}_{DY}}\ket{{\cal M}_{DY}}$. The values of the jet and soft
functions for the Drell-Yan process are given by
\begin{align}
\label{eq:DYJetSoft}
  {\cal J}^{(1,0)}_{DY} &= - \Lx\frac{1}{2\,\ep^2}
       + \frac{3}{4\,\ep}\Rx\Lx Q_q^2\, + Q_\ell^2\Rx\,,\qquad
   {\cal J}^{(0,1)}_{DY} = - \Lx\frac{1}{2\,\ep^2} + \frac{3}{4\,\ep}\Rx \CF\,,\nonumber\\
  {\cal J}^{(1,1)}_{DY} &= \Lx\frac{1}{4\,\ep^4} + \frac{3}{4\,\ep^3}
       + \frac{9}{16\,\ep^2}\Rx \CF\Lx Q_q^2\, + Q_\ell^2\Rx
       - \frac{1}{2\,\ep}\Lx\frac{3}{16} - \frac{3}{2}\zeta_2 +
       3\,\zeta_3\Rx \CF\,Q_q^2\,,\nonumber\\
  {\bf S}^{(1,0)}_{DY} &= -\frac{1}{2\,\ep}\LB \Lx Q_q^2 + Q_\ell^2\Rx
       \,\ln\Lx\frac{\mu^2}{-s}\Rx +
       2\,Q_q\,Q_\ell\Lx\ln\Lx\frac{\mu^2}{-t}\Rx -
       \ln\Lx\frac{\mu^2}{-u}\Rx\Rx\RB\,,\nonumber\\
  {\bf S}^{(0,1)}_{DY} &= -\frac{1}{2\,\ep}\CF\,\ln\Lx\frac{\mu^2}{-s}\Rx\,,\nonumber\\
  {\bf S}^{(1,1)}_{DY} &=
       \frac{1}{4\,\ep^2}\CF\,\ln\Lx\frac{\mu^2}{-s}\Rx
     \LB \Lx Q_q^2 + Q_\ell^2\Rx
       \,\ln\Lx\frac{\mu^2}{-s}\Rx +
       2\,Q_q\,Q_\ell\Lx\ln\Lx\frac{\mu^2}{-t}\Rx -
       \ln\Lx\frac{\mu^2}{-u}\Rx\Rx\RB\,.
\end{align}
We find complete agreement between our result and the expected
infrared structure presented in \eqn{eq:DYJetSoft}, including the
intrinsically \qcd$\times$\qed\ term in ${\cal J}^{(1,1)}_{DY}$.

\section{Results\label{sec:results}}
As our final result we present the interference of the finite hard-scattering
terms that appear in \eqn{eq:IRpoles}, defined by
\begin{equation}
2\*\aepi^2\mbox{Re}\left[\bra{H^{(1,0)}_{DY}}\ket{H^{(2,1)}_{DY}}+\bra{H^{(1,1)}_{DY}}\ket{H^{(2,0)}_{DY}} \right]=\nc\*Q_q^2\*Q_\ell^2\*e^4\*\CF\*B^{(1,1)}\,,
\end{equation}
where we performed the renormalization in the \MSbar scheme as described
in Section~\ref{sec:ultraviolet}. The infrared poles are subtracted
in $d$ dimensions with the help of \eqns{eq:IRpoles}{eq:DYJetSoft}.  We
decompose this mixed \qcd$\times$ \qed\ two-loop contribution with
respect to the charge factors
\begin{equation}
\label{eq:2loopdecomp}
B^{(1,1)}=
   Q_q\*Q_\ell\*B_{q\ell}^{(1,1)}
 + {t^2+u^2\over s^2}\*\left[
   Q_q^2\*B_{qq}^{(1,1)}
 + Q_\ell^2\*B_{\ell\ell}^{(1,1)}
 + \nc\*\sum_{q'} Q_{q'}^2\*B_{\Sigma q'}^{(1,1)}
 + \sum_{\ell'} Q_{\ell'}^2\*B_{\Sigma\ell'}^{(1,1)}
                   \right]\,,
\end{equation}
where the sum over $\ell'$ and $q'$ runs over all leptons and quark
flavors which are active in the closed fermion loop. Each of the five
terms corresponds to one of the classes of diagrams shown in
FIG.~\ref{fig:samplediagrams} (a)-($\mbox{d}$) and corresponds to a
gauge invariant subset of diagrams in this decomposition.  The
individual terms of \eqn{eq:2loopdecomp} are
\begin{eqnarray}
B_{qq}^{(1,1)}&=&
   {511\over 4} 
 - {83\over 3}\*\pi^2 
 + {67\over 30}\*\pi^4 
 - 60\*\z3 
 + \left(-93 + 10\*\pi^2 + 48\*\z3\right)\*\logsdmus
\nonumber\\
&+&\left(50 - {14\over 3}\*\pi^2\right)\*\logsdmusa{2} 
 - 12\*\logsdmusa{3}
 + 2\*\logsdmusa{4}
\,,\\
B_{\ell\ell}^{(1,1)}&=&
   128 
 - {112\over 3}\*\pi^2 
 + {49\over 18}\*\pi^4 
 + \left(14\*\pi^2-96\right)\*\logsdmus
\nonumber\\
&+&\left(50 - {14\over 3}\*\pi^2\right)\*\logsdmusa{2} 
 - 12\*\logsdmusa{3} 
 + 2\*\logsdmusa{4}
\,,\\
B_{\Sigma q'}^{(1,1)}&=&
   {155\over 9} 
 - {140\over 27}\*\pi^2 
 + 16\*\z3 
 + \left({28\over 9}\*\pi^2-{92\over 3}\right)\*\logsdmus
\nonumber\\
&+&{112\over 9}\*\logsdmusa{2}
 - {8\over 3}\*\logsdmusa{3}
\,,\\
B_{\Sigma\ell'}^{(1,1)}&=&
   {320\over 9} 
 - {140\over 27}\*\pi^2 
 + \left({28\over 9}\*\pi^2-{104\over 3}\right)\*\logsdmus
 + {112\over 9}\*\logsdmusa{2} 
 - {8\over 3}\*\logsdmusa{3}
\!\!,\,\\
B_{q\ell}^{(1,1)}&=&
    4\*{5\*u^2 - t^2\over s^2}\*\Limuds{4}
  - 4\*\left[{t\over s} + 4\*{u^2\over s^2}\*\logmuds\right]\*\Limuds{3}
\nonumber\\
&-& \Limuds{2}\*\left[
    {8\over 3}\*{t^2+u^2\over s^2}\*\pi^2 
  - 2\*{t\over s}\*\logmuds
  - 2\*{u\over s}\*\logmtds
\right.\nonumber\\&-&\left.
    {3\*u^2 + t^2\over s^2}\*\logmudsa{2}
  - {3\*t^2 + u^2\over s^2}\*\logmtdsa{2}
    \right]
 +  2\*\logsdmus\*\left[
    2\*\left({7\over3}\*{t^2+u^2\over s^2}\*\pi^2 
\right.\right.\nonumber\\&-&\left.\left.
               {19\*t^2 + 3\*t\*u + 16\*u^2\over s^2}
       \right)\*\logmuds 
  - 3\*{t - u\over s}\*\logmudsa{2}
                  \right]
\nonumber\\&+&
    2\*\logsdmusa{2}\*\left[
    {t - u\over s}\*\logmudsa{2}
  + 2\*{6\*u^2 + t\*u + 7\*t^2\over s^2}\*\logmuds
 \right]
\nonumber\\&-&
    8\*{t^2+u^2\over s^2}\*\logsdmusa{3}\*\logmuds
  + 4\*{t - u\over s}\*\z3\*\left[ 2\*\logmuds - 1 \right]
\nonumber\\&+&
    \pi^2\*\left[
    {4\over 3}\*{t - u\over s} 
 + {8\*u + 15\*t\over 3\*s}\*\logmuds 
  + {5\over 6}\*{u^2 + 3\*t^2\over s^2}\*\logmudsa{2}
       \right]
- \pi^4\*{2\over 15}\*{t - u\over s}
\nonumber\\&-&
    {3\*u^2 + t^2\over 6\*s^2}\*\logmudsa{4} 
  - {2\over 3}\*{t + 2\*u\over s}\*\logmudsa{3} 
  - 4\*{5\*u - 2\*t\over s}\*\logmudsa{2} 
\nonumber\\
 &-&40\*{t\over s}\*\logmuds 
  + {5\*t^2-u^2\over 3\*s^2}\*\logmtdsa{3}\*\logmuds 
  - \left(t\leftrightarrow u\right)
\,,
\end{eqnarray}
where $\mbox{Li}_n(z)=\sum_{k=1}^{\infty}{z^k\over k^n}$ is the
polylogarithm function and the symbol $(t\leftrightarrow u)$ stands
for the same terms a given before only with the Mandelstam
variables $u$ and $t$ interchanged.  As an additional check of our
calculation we have kept the complete dependence of the gauge parameter
in the gauge boson propagators and have verified their cancellation.

\section{Summary \& Conclusions\label{sec:summary}}
We have computed the two-loop virtual corrections to Drell-Yan
production at order $\alpha_s\,\alpha^3$.  The calculation of these
mixed \qcd$\times$\qed\ corrections includes two-loop corrections to the
quark vertex, one-loop corrections to the quark and lepton vertices,
vacuum polarization corrections to the photon propagator as well as
two-loop box diagrams connecting the hadronic and leptonic states.
The computation is accomplished by reducing all Feynman integrals to
a small set of master integrals. The latter ones are known analytically
to sufficiently high order in the $\ep$ expansion to allow us to derive an
analytical result for the finite amplitude. 

We have also shown that the infrared structure of the mixed amplitudes 
follows from the same universal factorization structure that governs
the pure QCD and QED amplitudes and we have determined the value of the
two-loop mixed anomalous dimension.\\

\paragraph*{Acknowledgments:}
We would like to thank Andreas Scharf and Doreen Wackeroth for useful
discussions.  This research was partially supported by the
U.~S.~Department of Energy under Contract No.~DE-AC02-98CH10886.

\appendix

\section{Bare next-to-leading order results in terms of master integrals\label{sec:BareRes}}
We present our bare results for the next-to-leading order processes in
terms of the master integrals and coefficients to all orders in $\ep$.
For the one-loop QCD and QED corrections we adopt the decomposition of
the squared matrix element as given in \eqn{eq:decomp}; all 
quantities are considered as bare. We find
\begin{eqnarray}
\label{eq:Ab01}
A_{B}^{(0,1)}&=&
{A^{(0,0)}\over4}\*\left(1-{2\over\ep}-2\*\ep\right)\*B^r_{0}(s)\,,\\
\label{eq:Ab10}
A_{B}^{(1,0)}&=&
   (Q_q^2+Q_\ell^2)\*A_{B}^{(0,1)}
 + \left(\sum_{\ell'}Q_{\ell'}^2+\nc\*\sum_{q'}Q_{q'}^2\right)\*
 A^{(0,0)}\*{1-\ep\over 2\*\ep-3}\*B^r_0(s)
\nonumber\\
&+&Q_q\*Q_\ell\*\Bigg[
   \left(10 - {4\over\ep} - 2\*\ep\right)\*{t-u\over s}\*B^r_0(s)
 + \left(6 - {4\over\ep} + 2\*\ep\right)\*{t-u\over s}\*B^r_0(u) 
\nonumber\\
& &\qquad\quad+\left( 2\*{u\*(t^2 + 3\*u^2)\over s} - 3\*\ep\*s\*u\right)\*D^r_0(s,u)
 - \left(t\leftrightarrow u\right)
   \Bigg]\,,
\end{eqnarray}
with the integrals
\begin{equation}
B^r_0(s)=\left(4\*\pi\*\mu^2\right)^{\ep}\*e^{-\ep\gamma_E}\*2\*\mbox{Re}\left[I^{(1)}_2(s)\right]\quad\mbox{and}\quad
D^r_0(s,u)=\left(4\*\pi\*\mu^2\right)^{\ep}\*e^{-\ep\gamma_E}\*2\*\mbox{Re}\left[I^{(1)}_4(s,u)\right]\,,
\end{equation}
where $e\simeq2.71828$ is again Euler's number and
$\gamma_E\simeq0.577216$ is the Euler-Mascheroni constant. The values
of the master integrals $I^{(1)}_2$ and $I^{(1)}_4$ are given in
Appendix~\ref{sec:MI}.  In the coefficients of the master integrals of
\eqns{eq:Ab01}{eq:Ab10}, spurious poles in $\ep$ appear, which arise
while solving the linear system of IBP equations. As a result, one
must know the master integrals which are multiplied by such spurious poles
at higher order in the $\ep$ expansion. The same situation also occurs in the
two-loop amplitude. In principle, those spurious poles could be
avoided by choosing an epsilon finite
basis~\cite{Chetyrkin:2006dh}. However, since all necessary master
integrals are known either in closed form or to sufficiently high
order in $\ep$, we retain the standard basis of master integrals.

\section{Master Integrals\label{sec:MI}}
The reduction process relates complicated integrals with many terms in
the numerators and denominators to ``simpler'' integrals with fewer
terms in both numerators and denominators.  In general, it is
preferred that the master integrals have numerators equal to unity,
and denominators which only contain propagators of unit strength, but
this preference cannot always be satisfied.

In this calculation, we encounter eighteen two-loop master
integrals. Of these, eight represent distinct topologies which are
shown in FIG.~\ref{fig:MI}; the others are related to these eight by
relabeling the external legs.  Only one of the distinct topologies has
an irreducible numerator (or, equivalently, a doubled propagator in
the denominator).

All of the master integrals needed for this calculation are known in the
literature.  The double box integrals, FIG.~\ref{fig:MI}($\mbox{g}_1$)
and FIG.~\ref{fig:MI}($\mbox{g}_2$) are known as Laurent expansions in
the dimensional regularization parameter $\ep$.  The others are all
known in closed form and can be readily computed using standard Feynman
parametrization techniques. 

In the following we define the master integrals $I^{(k)}_{p;s}$ with
loop momenta $k_1$ and $k_2$ in Minkowski space, where the superscript
$k$ indicates the number of loops, the subscript $p$ denotes the number
of propagators and $s$ enumerates integrals with the same number of
loops and propagators.  For clarity, we also indicate the Mandelstam
variables that appear as arguments.

\subsection{One-loop master integrals}
At one-loop order we have the five master integrals $I^{(1)}_2(s)$,
$I^{(1)}_2(t)$, $I^{(1)}_2(u)$, $I^{(1)}_4(s,t)$ and $I^{(1)}_4(s,u)$, which
are define by
\begin{equation}
I^{(1)}_2(s)=e^{\ep\gamma_E}\*\int{d^dk_1\over i\*\pi^{d/2}}\*{1\over D_1\*D_3}\,,\qquad
I^{(1)}_4(s,u)=e^{\ep\gamma_E}\*\int{d^dk_1\over i\*\pi^{d/2}}\*{1\over D_1\*D_2\*D_3\*D_4}\,,
\end{equation}
with
\[
\begin{array}{ll}
D_1=k_1^2+i\*\dep,          & D_2=(k_1-p_1)^2+i\*\dep,\\
D_3=(k_1-p_1-p_2)^2+i\*\dep,& D_4=(k_1-p_1-p_2-p_3)^2+i\*\dep.
\end{array}
\]
Their values are
\begin{eqnarray}
\label{eq:I12s}
I^{(1)}_2(s)&=&e^{\ep\*\gammaE}\*(-s)^{-\ep}\*
             {\Gammap{1-\ep}{2}\*\Gamman{\ep}\over\Gamman{2-2\*\ep}}\,,\\
I^{(1)}_4(s,t)&=&{2\*e^{\ep\*\gammaE}\over s\*t}\*
                {\Gamman{1+\ep}\*\Gammap{-\ep}{2}\over \Gamman{1-2\*\ep}}\*\left[ 
                (-s)^{-\ep}\*\hypgeo{1}{-\ep}{1-\ep}{1+{s\over t}}
\right.\nonumber\\&&\qquad\qquad\qquad\qquad\quad\;\left.
               +(-t)^{-\ep}\*\hypgeo{1}{-\ep}{1-\ep}{1+{t\over s}}
              \right]\,,
\end{eqnarray}
where $\hypgeo{a}{b}{c}{z}=\sum_{k=0}^{\infty}(a)_k (b)_k/ (c)_k
z^k/k!$ are hypergeometric functions, $(a)_n=\Gamma(a+n)/\Gamma(a)$ is
the Pochhammer symbol and $\Gamma(x)$ is the gamma function.

\subsection{Two-loop master integrals}
\subsubsection{Three-Line Topologies}
There is one distinct three-line topology, shown in
Fig.~\ref{fig:MI}(a), which we label $\MIi$ 
and define by
\begin{equation}
\MIi=e^{2\*\ep\gamma_E}\*\int{d^dk_1\over i\*\pi^{d/2}}\*{d^dk_2\over i\*\pi^{d/2}}\*
     {1\over D_5\*D_6\*D_7}\,,
\end{equation}
with
\[
\begin{array}{lll}
D_5=(k_1-p_2)^2+i\*\dep\,\quad, & D_6=(k_1-k_2)^2+i\*\dep\,\quad, & D_7=(k_2+p_1)^2+i\*\dep.
\end{array}
\]
Its value is
\begin{eqnarray}
\MIi&=& e^{2\*\ep\*\gammaE}\*(-s)^{1-2\*\ep}\*
       {\ep^3\*\Gammap{-\ep}{3}\*\Gamman{-1+2\*\ep}\over\Gamman{3 - 3\*\ep}}\,.
\label{eqn:SunsetSb}
\end{eqnarray}
In addition, we also need $\MIiii$ and $\MIii$.

\subsubsection{Four-Line Topologies}
There are four four-line master integrals, $\MIiv$, $\MIv$, $\MIvii$
and $\MIvi$ with two distinct four-line topologies. One is a simple
iterated bubble diagram, shown in Fig.~\ref{fig:MI}(b), which
evaluates to the square of the expression in \eqn{eq:I12s}; the other is
shown in Fig.~\ref{fig:MI}(c). They are defined by
\begin{equation}
\MIiv\ =e^{2\*\ep\gamma_E}\*\int{d^dk_1\over i\*\pi^{d/2}}\*{d^dk_2\over i\*\pi^{d/2}}\*
        {1\over D_1\*D_3\*D_8\*D_9}\,,\quad
\MIv\ =e^{2\*\ep\gamma_E}\*\int{d^dk_1\over i\*\pi^{d/2}}\*{d^dk_2\over i\*\pi^{d/2}}\*
       {1\over D_5\*D_6\*D_8\*D_{10}}
\end{equation}
with
\[
D_8=k_2^2+i\*\dep,\quad
D_9=(k_2-p_1-p_2)^2+i\*\dep,\quad 
D_{10}=(k_1+p_1)^2+i\*\dep,
\]
and are given by $\MIiv=(I^{(1)}_2(s))^2$,
\begin{eqnarray}
\MIv\ &=&e^{2\*\ep\*\gammaE}\*(-s)^{-2\*\ep}\*
      {\Gamman{1-2\*\ep}\*\Gammap{-\ep}{2}\*\Gamman{1+\ep}\*\Gamman{1+2\*\ep}\over 
      2\*(1-2\*\ep)\*\Gamman{2-3\*\ep}}\,.
\label{eqn:TriSa}
\end{eqnarray}

\subsubsection{Five-line Topologies}
There are six five-line master integrals $\MIx$, $\MIix$, $\MIviii$,
$\MIxi$, $\MIxiii$ and $\MIxii$ with two distinct five-line topologies,
which are shown in Fig.~\ref{fig:MI}(d) and Fig.~\ref{fig:MI}(e). The
first topology has a bubble connecting two adjacent corners of a box,
the other five-line topology is a box diagram with a diagonal-line
connecting opposite corners. They are defined by
\begin{eqnarray}
\MIviii\ &=&e^{2\*\ep\gamma_E}\*\int{d^dk_1\over i\*\pi^{d/2}}\*{d^dk_2\over i\*\pi^{d/2}}\*
            {1\over D_1\*D_3\*D_6\*D_{11}\*D_{12}}\,,\\
\MIxii\  &=& e^{2\*\ep\gamma_E}\*\int{d^dk_1\over i\*\pi^{d/2}}\*{d^dk_2\over i\*\pi^{d/2}}\*
            {1\over D_3\*D_6\*D_8\*D_{11}\*D_{12}}\,,
\end{eqnarray}
with
\[
D_{11}=(k_1+p_3)^2+i\*\dep,\quad D_{12}=(k_2-p_2)^2+i\*\dep\,,
\]
Their results read
\begin{eqnarray}
\MIviii&=& 
-{e^{2\*\ep\*\gammaE}\over s}\*{\Gammap{-\ep}{2}\*\Gamman{-1+2\*\ep}\over\Gamman{1-3\*\ep}}\*\bigg[
  (-u)^{-2\*\ep}\*\Gamman{1-\ep}\*\hypgeo{1}{-\ep}{1-\ep}{-{t\over s}}
\nonumber\\&+&
  (-s)^{-2\*\ep}\*\Gamman{1+\ep}\*\Gamman{1-2\*\ep}\*\hypgeo{1}{\ep}{1-\ep}{-{t\over s}}
   \bigg]\,,\\
\MIxii\ &=& 
e^{2\*\ep\*\gammaE}\*{\Gammap{-\ep}{3}\*\Gamman{2\*\ep}\over 2\*t\*\Gamman{1-3\*\ep}}\*
  \Bigg[
  (-u)^{-2\*\ep}\*\left(1-\hypgeo{1}{-2\*\ep}{1-2\*\ep}{-{t\over s}}\right)
\nonumber\\&+&
  (-s)^{-2\*\ep}\*\left(1-\hypgeo{1}{-2\*\ep}{1-2\*\ep}{-{t\over u}}\right)
  \Bigg]\,.
\label{eqn:XSTBoxa}
\end{eqnarray}

\subsubsection{Six-line Topologies}
There is only one six-line master integral, the non-planar triangle diagram,
shown in Fig.~\ref{fig:MI}(f) and defined by
\begin{equation*}
\MIxiv\ = e^{2\*\ep\gamma_E}\*\int{d^dk_1\over i\*\pi^{d/2}}\*{d^dk_2\over i\*\pi^{d/2}}\*
          {1\over D_1\*D_6\*D_7\*D_{10}\*D_{12}\*D_{13}}\,,
\quad\mbox{with}\quad D_{13}=(k_2-k_1-p_2)^2+i\*\dep.
\end{equation*}
Its result can be expressed with the help of generalized hypergeometric functions
$\mbox{}_{p}F_q(a_1,\dots,a_p;b_1,\dots,b_q;z)=\sum_{k=0}^{\infty}
(a_1)_k\dots(a_p)_k/((b_1)_k\dots(b_q)_k)z^k/k!$ by
\begin{eqnarray}
\MIxiv\ &=&
-e^{2\*\ep\*\gammaE}\*(-s)^{-2-2\*\ep}\*\Gamman{1+2\*\ep}\*\Bigg[
\nonumber\\&& 
 - {\Gamman{1-\ep}\*\Gammap{1-2\*\ep}{4}\*\Gamman{1+\ep}\*\Gammap{1+2\*\ep}{2}\over
    \ep^4\*\Gammap{1-4\*\ep}{2}\*\Gamman{1+4\*\ep}}
 - {4\*\Gammap{1-\ep}{2}\*\Gamman{1-2\*\ep}\over\ep^4\*\Gamman{1-4\*\ep}}
\nonumber\\&& 
 + {\Gammap{1-\ep}{2}\*\Gamman{1-2\*\ep}\*\Gamman{1+\ep}\over
    2\*\ep^4\*\Gamman{1-3\*\ep}}\*
   \ghypgeo{1}{-2\*\ep}{-4\*\ep}{1-2\*\ep}{1-3\*\ep}{1}
\nonumber\\&& 
 + {4\*\Gammap{1-\ep}{2}\*\Gamman{1-2\*\ep}\*\Gamman{1+\ep}\*\Gamman{1+2\*\ep}\over
    \ep^4\*\Gamman{1-4\*\ep}\*\Gamman{1+3\*\ep}}\*
     \ghypgeo{\ep}{\ep}{1+2\*\ep}{1+\ep}{1+3\*\ep}{1}\,
\nonumber\\&&
 - {\Gammap{1-\ep}{3} \over 2\*\ep^4\*\Gamman{1-3\*\ep}}\*
   \gghypgeo{1}{1-\ep}{-2\*\ep}{-4\*\ep}{1-2\*\ep}{1-2\*\ep}{1-3\*\ep}{1}
 \Bigg]\,.
\label{eqn:DTriNPb}
\end{eqnarray}
Note that the closed-form expression given above appears to differ
slightly from that given by Ref.~\cite{Gehrmann:2005pd}.  However, by
rearranging the $\Gamma$-functions and applying various hypergeometric
identities, one finds that the two expressions are exactly equal.

\subsubsection{Seven-line Topologies}
There are four seven-line master integrals $\MIxvi$, $\MIxv$, $\MIxviii$
and  $\MIxvii$ with two distinct topologies. One is the double-box
topology where all propagators are of unit strength, shown in
FIG.~\ref{fig:MI}($\mbox{g}_1$). It is defined by 
\begin{equation}
\MIxv\ =  e^{2\*\ep\*\gammaE}\*\int{d^dk_1\over i\*\pi^{d/2}}\*{d^dk_2\over i\*\pi^{d/2}}\*
          {1\over D_1\*D_3\*D_5\*D_6\*D_8\*D_9\*D_{14}}\,,
\quad\mbox{with}\quad D_{14}=(k_2+p_3)^2+i\*\dep\,,
\end{equation}
and known as a Laurent expansion in $\ep$~\cite{Smirnov:1999gc}   
\begin{eqnarray}
\MIxv&=& 
  -{(-s)^{-2-2\ep}\over u}\*\bigg\{
 - {4\over\ep^4} 
 + 5\*{\ell\over\ep^3} 
 - {1\over\ep^2}\*\left[
   2\*\ell^2 
 - 15\*\z2     \right]
 - {1\over\ep}\*\bigg[
   4\*\nLi{3}{-x}
 - 4\*\ell\*\nLi{2}{-x}
\nonumber\\
&+&2\*\nLi{1}{-x}\*\left(
   \ell^2 
 + 6\*\z2\right)
 + {2\over3}\*\ell^3 
 + 33\*\z2\*\ell
 - {65\over3}\*\z3
              \bigg]
 + {4\over3}\*\ell^4 
 + 36\*\z2\*\ell^2
 - {88\over3}\*\z3\*\ell 
\nonumber\\
&+&87\*\z4
 - 4\*\left(\Sx{2}{2}{-x}
 - \ell\*\Sx{1}{2}{-x}\right)
 + 44\*\nLi{4}{-x}
 + 4\*\nLi{3}{-x}\*\left(
   \nLi{1}{-x}
 - 6\*\ell\right)
\nonumber\\
&+&2\*\nLi{2}{-x}\*\left(
   \ell^2
 - 2\*\ell\*\nLi{1}{-x}
 + 20\*\z2\right)
 + \pLi{1}{-x}{2}\*\left(
   \ell^2
 + 6\*\z2\right)
\nonumber\\
&+&{2\over3}\*\nLi{1}{-x}\*\left(
   4\*\ell^3
 + 30\*\z2\*\ell
 - 6\*\z3\right)
 + {\cal O}\Lx\ep\Rx\bigg\}\,,
\label{eqn:DBox}
\end{eqnarray}
with $\ell=\nlog{x}$, $x =u/s$ and the generalized polylogarithm function 
$\Sx{n}{p}{z}={(-1)^{n+p-1}/(n-1)!/p!}\int_{0}^{1}dt'\log^{n-1}{(t')}\*\log^p{(1-z\*t')}/t'$.

There are two equivalent representations for the second seven-line
topology. One is the double-box with a doubled propagator. The other
representation is a double-box with an irreducible numerator, shown in
FIG.~\ref{fig:MI}($\mbox{g}_2$). The latter is defined by
\begin{equation}
\MIxvii\ = e^{2\*\ep\gamma_E}\*\int{d^dk_1\over i\*\pi^{d/2}}\*{d^dk_2\over i\*\pi^{d/2}}\*
           {(k_1+p_3)^2\over D_1\*D_3\*D_5\*D_6\*D_8\*D_9\*D_{14}}\,.
\end{equation}
When one uses the integral with the doubled propagator, the reduction
procedure generates a spurious pole in $\ep$, meaning that one needs
the double-box integrals expanded to order $\ep^{1}$.  When one instead uses
the above double-box with an irreducible numerator, the reduction does not
generate the extra pole, meaning that one only needs to expand the
integrals to order $\ep^0$.

The double-box with an irreducible numerator was first calculated in
\cite{Anastasiou:2000kp}, with the result
\begin{eqnarray}
\MIxvii&=&
   e^{2\*\ep\*\gamma_E}\*(-s)^{-2-2\ep}\*\Gammap{1+\ep}{2}\*\Bigg\{
   {9\over4\*\ep^4} 
 - {2\over\ep^3}\*\ell 
 - {14\*\z2\over\ep^2}
 + {1\over\ep}\*\left[ 
   {4\over3}\*\ell^3 
 + 28\*\z2\*\ell 
\right.\nonumber\\
&+&\left.
   4\*(\ell^2 + 6\*\z2)\*\nLi{1}{-x}
 + 8\*\nLi{3}{-x}
 - 8\*\ell\*\nLi{2}{-x }
 - 16\*\z3 \right]
 - {4\over3}\*\ell^4 
 - 26\*\z2\*\ell^2 
\nonumber\\
&-&\left[{16\over 3}\*\ell^3 + 52\*\z2\*\ell\right]\*\nLi{1}{-x}
 - 5\*\left[\ell^2 + 6\*\z2\right]\*\pLi{1}{-x}{2}
 + \left[
   6\*\ell^2
 + 20\*\ell\*\nLi{1}{-x}
\right.\nonumber\\
&-&\left.
   8\*\z2\right]\*\nLi{2}{-x}
 + \left[8\*\ell - 20\*\nLi{1}{-x}\right]\*\nLi{3}{-x}
 + 20\*\Sx{2}{2}{-x} 
 - 20\*\ell\*\Sx{1}{2}{-x}
\nonumber\\
&-&28\*\nLi{4}{-x}
 + \left[
   28\*\ell 
 + 20\*\nLi{1}{-x}
   \right]\*\z3 
 - 14\*\z4
 + {\cal O}\Lx\ep\Rx
 \Bigg\}\,.
\label{eqn:DBoxIrr}
\end{eqnarray}

\bibliographystyle{apsrev}

\end{document}